\documentclass[printer,floatfix]{aa}
\usepackage{graphicx}

\begin{document}
\def\al{&\!\!\!\!}
\def\x{{\bf x}}
\def\f{\frac}

\title{The distinguishing factor for gravity models: stellar population synthesis }

\author{ Akram Hasani Zonoozi,
         \inst{} 
         \thanks{\email{a.hasani@iasbs.ac.ir}}
          \and
         Hosein Haghi
        \inst{}          }
\offprints{A. Hasani Zonoozi}

 \institute{ Institute for Advanced Studies in Basic Sciences (IASBS), P. O. Box 45195-1159,
    Zanjan, Iran }

\date{Received *****; Accepted *****}

\abstract{ Alternative gravitations of Milgrom (MOND), Moffat (MOG),
and  CDM scenarios all simulate rotation curves of spirals with
reasonable details. They display significant disparities however in
predicting the stellar mass-to-light ($M_*/L$) ratios of the
galaxies. We maintain this feature could serve as a distinguishing
factor between different alternative theories. We analyze the
rotation curves of 46 low- and high-surface brightness galaxies and
compare the resulting $M_*/L$s with the predictions of the Stellar
Population Synthesis (SPS) scheme. The color-$M_*/L$ correlation
obtained for MOND is consistent with predictions of SPS models. MOG
does not show this consistency, and the $M_*/L$s of CDM model shows
large dispersions. Furthermore, $M_*/L$ ratios of MOND with
Bekenstein interpolating function favor Kroupa's initial mass
function (IMF) of the SPS scheme, while those of MOND with standard
and simple interpolating functions are consistent with Salpeter's
IMF. Here is another indication to differentiate between different
IMFs that are used in SPS context. }
\authorrunning{Hasani et al.}
\titlerunning{SPS scheme, a discriminant between gravity models}
\maketitle

\section{Introduction}

The gravitational force of the observable mass of large astronomical
systems, galaxies, clusters of galaxies, or for that matter, the
universe in general, is not sufficiently strong to explain the
observed dynamics of the systems. To resolve the dilemma, one main
school of investigators has resorted to dark matter/dark energy
scenarios. In spite of extensive efforts, however, no one has so far
reported a direct identification of the hypothesized dark entity
through non-gravitational interactions with the observable matter.
This lack of direct identification has inspired an equally intensive
effort to contemplate alternative theories of gravitation. The
Modified Newtonian Dynamics (MOND) of Milgrom (1983) and of
Bekenstein (2004), the Modified Gravity (MOG) of Moffat (2005), the
Nonlocal Nonlinear gravity of Sobouti (2008a, b, 2009), and
varieties of $f(R)$ gravities (\cite{cap02}, \cite{cap0607},
\cite{car04}, \cite{sob07}, \cite{sob.etal09}) fall in this
category.

Rotation curves of spiral galaxies as measured by the 21 cm line of
HI often extend well beyond the optical disks of the galaxies and
provide a valuable body of data to determine the radial dependency
of the gravitational forces in galactic scales. In this paper we
construct rotation curves of a large sample of galaxies from the
distribution of their detectable matter through three different
gravity models, MOND, MOG, and Newtonian gravity plus cold dark
matter (CDM) halos. At first glance all three models seem to
reproduce the observed data with reasonable detail. On a deeper
examination, however, we find significant disparities in their
predictions of stellar mass-to-light, $M_*/L$, ratios. To
differentiate between the models we resort to stellar population
synthesis, SPS, analysis and the color-$M_*/L$ correlation predicted
therein through various initial mass functions (IMF). There is the
possibility to use this feature to discriminate between different
gravity models and different IMFs.

The paper is organized as follows: In Sect. 2 we give a brief review
of the different gravity models used in our analysis. In Sect. 3 we
describe  our galaxy sample. Fits to the observed rotation curves
are discussed in Sect. 4. Numerical results and brief concluding
remarks are given in Sects. 5 and 6.

\section{Alternative gravity models}\label{model}

In this section we review the basic tenets of two alternative
gravities as well as the Newtonian gravity plus CDM halos. We
present the end formulas that we will use in the study of the
dynamics of galaxies. All three accommodate the two main asymptotic
features of the rotation curves of spirals: the slow non-Keplerian
decline of the curves at large distances from the galaxy, and the
Tully-Fisher (TF) relation, which is the approximate proportionality
of the asymptotic speed of an orbiting object to the fourth root of
the mass of the galaxy (\cite{tul77}).

Actually there is much debate on how fast or slow the rotation
curves decline, if at all (\cite{per96}, \cite{sal07},
\cite{gen08}). There are also refinements and redefinitions to the
Tully-Fisher relation. McGaugh (2005) prefers to use the total
baryonic (stellar + gaseous) mass in TF relation to accommodate the
gas-rich galaxies. See also Stark et al. (2009), for a calibration
of the baryonic TF relation with the help of gas dominated galaxies.
Nonetheless, both assumptions are adequate approximations to the
observed data and will be employed in this paper.

\subsection{Modified Newtonian Dynamics of Milgrom, MOND }

Based on observations of galactic rotation curves, Milgrom (1983)
argues that the Newtonian dynamics is not viable below a certain
universal acceleration,  $a_{0}\simeq1.2\times10^{-10} \rm{m
/sec}^{2}$. To comply with the Tully-Fisher relation he modifies the
law of motion to have an asymptotic acceleration proportional to the
square root of the Newtonian acceleration.  The MOND acceleration,
$g_{mond}$, and the Newtonian one, $g_N$, are connected through Eq.
(\ref{gmond}) below
\begin{equation}\label{gmond}
\frac{g_{mond}}{a_0}~\mu \left(\frac{g_{mond}}{a_0}\right)=
\frac{g_N}{a_0},
\end{equation}
where $\mu(x)$ is an interpolating function for transition from the
Newtonian to the MONDian regime.  It runs smoothly from $\mu(x)=x$
for $x<<1$ to $\mu(x)=1$ for $x>>1$. Here we adopt three functions
commonly used, the standard interpolating function of Bekenstein \&
Milgrom (1984):
\begin{equation}\label{standard}
\mu_{1}(x)=\frac{x}{(1+x^2)^{1/2}},
\end{equation}
the simpler function of Famaey \& Binney (2005),
\begin{equation}\label{simple}
\mu_2(x)=\frac{x}{1+x},
\end{equation}
and Bekenstein's interpolating function (\cite{bek04})
\begin{eqnarray}\label{mond3}
\mu_3(x)=\frac{(1+4x)^{1/2}-1}{(1+4x)^{1/2}+1}.
\end{eqnarray}
Hereafter, the analysis using $\mu_1$, $\mu_2$ and $\mu_3$ will be
refereed to as MOND1, MOND2 and MOND3, respectively.

\subsection{Modified Gravity of Moffat, MOG}

Modified Gravity of Moffat consists of three theories of gravity:
the nonsymmetric gravity theory (NGT), the metric-skew-tensor
gravity (MSTG) theory, and the scalar-tensor-vector gravity (STVG).
They rely on the existence of a massive vector field universally
coupled to matter. Moffat maintains that MOG explains the rotation
curves of galaxies, clusters of galaxies, and cosmological issues
without resorting to dark matter
(\cite{Mof94,mof05,Mof06a,moffat09c}). Good fits to astrophysical
and cosmological data have been obtained with his recent version of
STVG. One notable feature of NGT, MSTG, and STVG is that the
modified acceleration at weak gravitational fields has a Yukawa-type
addition to the Newtonian acceleration. In the weak field limit,
STVG, NGT, and MSTG produce similar results. The recipe for the
gravitational force of a spherically distributed mass, $M(r)$, is
(\cite{Mof06a})
\begin{eqnarray}
\al\al g_{mog}=\frac{G(r)M(r)}{r^2},\label{gmog}\\
\al\al G(r)= G_N
\times\left\{1+\alpha(r)\left[1-e^{-r/r_0}\left(1+\frac{r}{r_0}\right)\right]\right\}, \nonumber
\end{eqnarray}
where  $G_N$ is the Newtonian gravitational constant, $M(r)$ is the
baryonic mass inside the radius $r$, and $\alpha(r)=[{M_0/M(r)}]^{1/2}$.
The parameters  $M_0$ or $r_0$ determine the
coupling strength of the vector field to the baryonic matter and to
the range of the force, respectively. They are not universal
constants and vary with the  size of the systems (\cite{brow06,
hagh10}). In galactic scales, they are determined by analyzing the
best fit  of the theory to the  rotation curves of LSB and HSB
galaxies. For normal size galaxies, they are reported as
$M_0=9.6\times10^{11} M_\odot$ and $r_0=13.9$ kpc, and for dwarf
galaxies, as $M_0=2.4\times10^{11} M_\odot$ and $r_0=9.7$ kpc
(\cite{Mof06a}). An empirical fitting of $M_0$ versus $r_0$ for a
wide range of spherically symmetric systems, from  solar size
 to clusters of galaxies is depicted in Fig. 2 of Brownstein \&
 Moffat (2006). The MOG gravitation tends to the Newtonian one as
$M_0\rightarrow 0$ and $r_0\rightarrow\infty$.

\subsection{Newtonian Gravity plus Cold Dark Matter, CDM }

In this scenario, gravitation is Newtonian.  To account for the
nonclassical behavior of the  rotation curves one adds a spherically
symmetric dark halo to the galaxy. Here, we consider a NFW halo with
the density distribution
$$\rho_{NFW}(r)= \frac{\rho_s}{(r/r_s)(1+r/r_s)^2},$$
and the gravitational acceleration
\begin{equation}
g_{NFW}=4\pi G \rho_s
r_s(\frac{r_s}{r})^2\left[\ln\left(1+\frac{r}{r_s}\right)-\frac{r/r_s}{(1+r/r_s)}\right],\label{cdm}
\end{equation}
\noindent where $r_s$ and $\rho_s$ are the characteristic radius and
density of the distribution (\cite{nfw}). The NFW density comes from
numerical simulations of $\Lambda$CDM theory in the framework of
structure formation. There, one also finds that these parameters are
correlated to each other as in Eqs. (\ref{rhos} and \ref{c}) below,
leaving only one free parameter to characterize the halo (see
\cite{bul01,wec02}, and \cite{net07} for details).  Thus,

\begin{equation} \rho_s=
\frac{\Delta}{3}\frac{c^3}{ln(1+c)-c/(1+c)}\rho_c,\label{rhos}
\end{equation}

\begin{equation} c= 13.6
\left(\frac{M_{vir}}{10^{11}M_{\odot}}\right)^{-0.13},
r_s=8.8\left(\frac{M_{vir}}{10^{11}M_{\odot}}\right)^{0.46} kpc,\label{c}
\end{equation}

\noindent where $\rho_c$ is the critical density of the Universe and
$\Delta=200$ is the virial overdensity at redshift z=0
(\cite{bra98}).\\

\section{Observational data}\label{sample}

\begin{figure*}{}
\resizebox{18cm}{!}{\includegraphics[width=70mm,height=80mm]{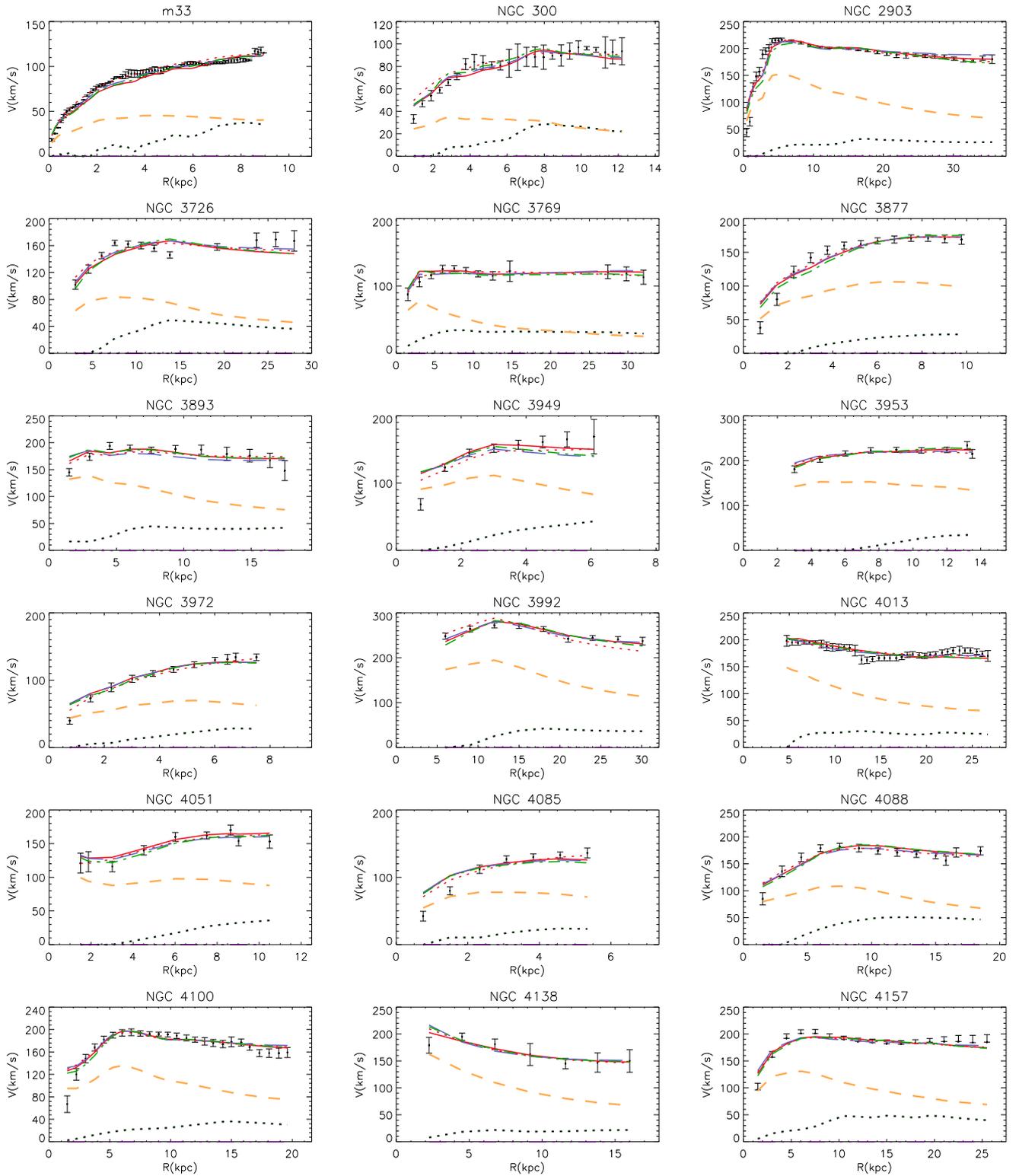}}
\caption{ Rotation curves of 32 mainly HSB galaxies. The points with
vertical error bars are the observed data. The dotted (black) and
short dashed lines are contributions of the gaseous and stellar
components to the Newtonian rotation speeds, respectively. The long
dashed (blue) line is that of MOND1. The solid line is the rotation
curve constructed through MOND3. The $M_*/L$s of MOND3 are used in
plotting the stellar component. The dashed-dotted (green) and dotted
(red) lines are those of the MOG and NFW models, respectively.}
\label{f1}
\end{figure*}

\begin{figure*}{}
\resizebox{18cm}{!}{\includegraphics[width=75mm,height=65mm]{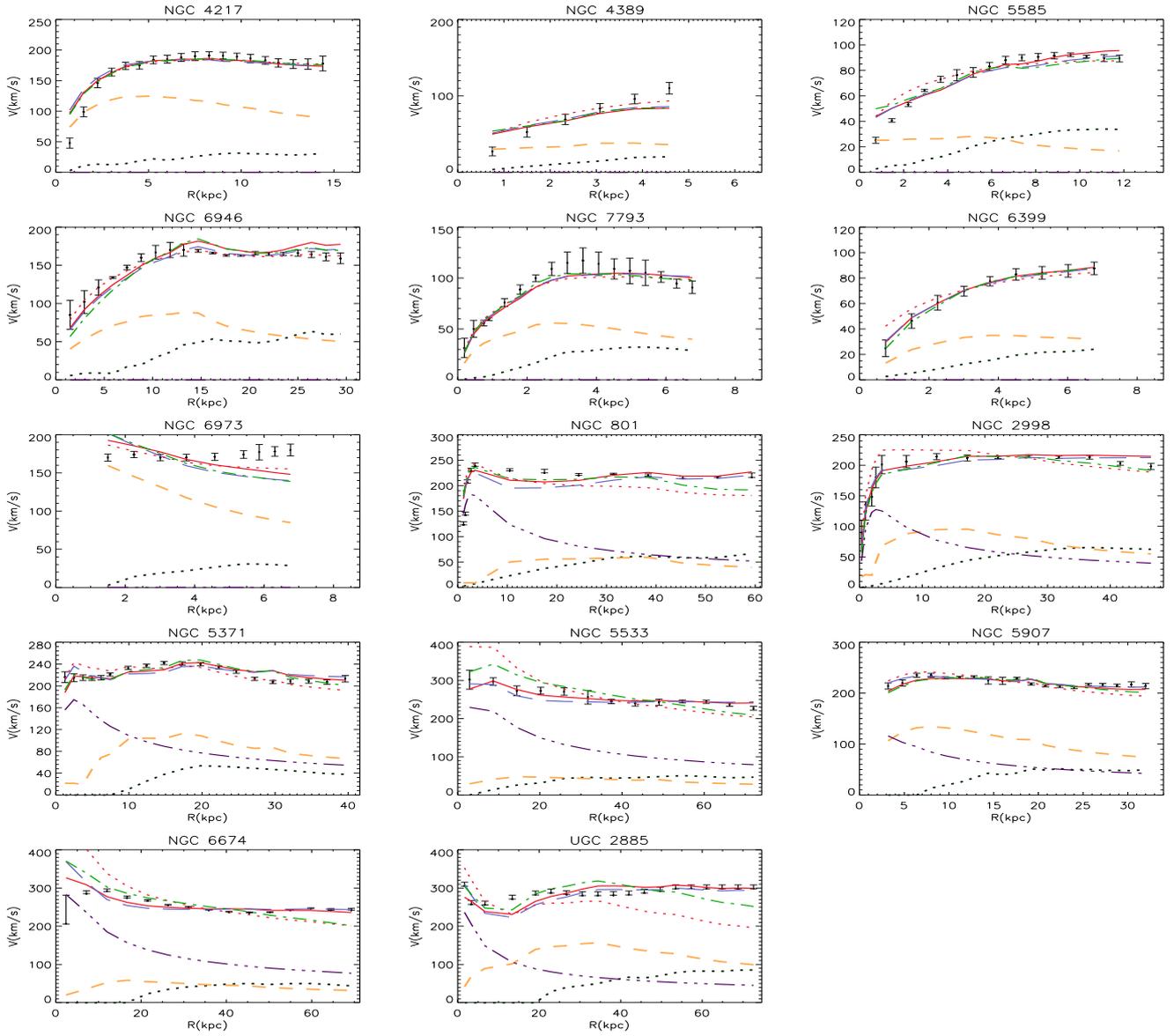}}
\caption{ Fig. 1 continued. The last seven galaxies have a bulge
component, depicted as dashed-dotted lines.  }\nonumber \label{f2}
\end{figure*}

\begin{figure*}{}
\begin{center}
\resizebox{18cm}{!}{\includegraphics[width=70mm,height=67mm]{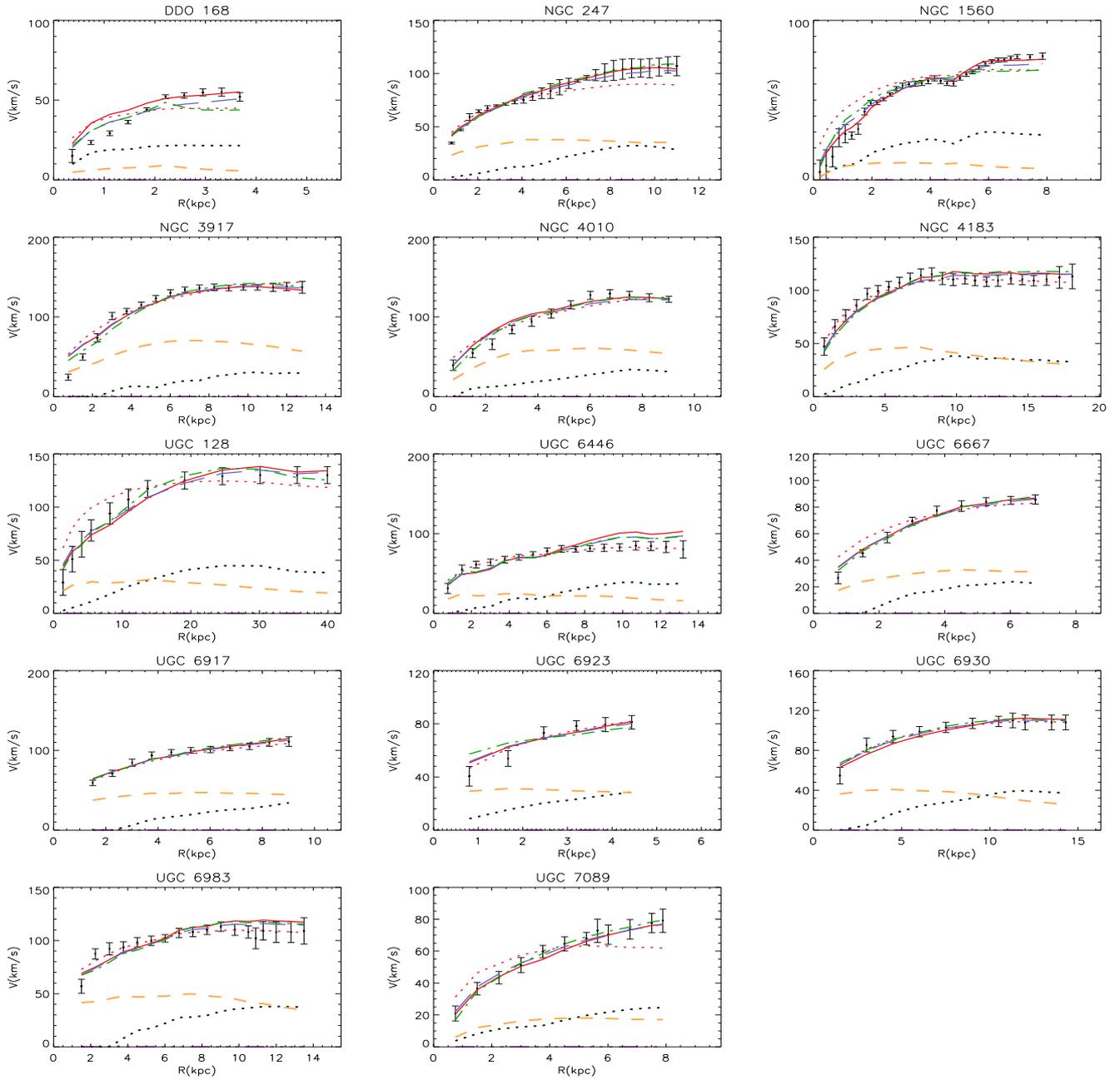}}
\caption{Rotation curves of 14 LSB galaxies.  Legend as in Fig. 1.
}\label{f3}
\end{center}
\end{figure*}

There are diverse morphological types of galaxies with diverse
shapes and sizes to their rotation curves.  Our sample, a collection
of 46 galaxies taken from  Sanders (1996), McGaugh \& de Blok
(1998), Sanders \& Verheijen (1998), and Begeman (1991),
accommodates these diversities. Members of the  sample have well
measured rotational speeds and accurate surface photometry. They are
listed in Table \ref{t1} and shown in Figs. \ref{f1} - \ref{f3}.

The sample includes several very large and luminous members with
well-extended rotation curves, e.g., UGC 2885, NGC 801, and NGC
2903. They have high surface brightness (HSB), massive stellar
component, and low gas content. Typically, their rotation curve
rises steeply to a maximum and declines slowly into an almost
horizontal asymptote.   There are also a number of dwarf,
gas-dominated, and low-surface brightness (LSB) galaxies, e.g. DDO
168. There is no conspicuous maximum, and in some galaxies not even
a flat asymptote to their rotation curve. It is generally believed
that deviations from the classical dynamics is  more pronounced in
LSBs than in HSBs. (\cite{mcg98a,san07,gen10})

Twenty-eight members of the sample, of both HSB and LSB types, are
located in the Ursa Major cluster of galaxies, believed to be at the
distance of about 15.5 Mpc (\cite{tul97}). Seven of the galaxies,
listed in Table \ref{t3}, have central bulges and are treated
differently, the reason is explained below. For a full description
of the sample the interested reader is referred to Sanders and
McGaugh (2002).

\section{Constructing rotation curves}\label{rc}

We calculate the rotation speed of a test object circling the galaxy
as a function of distance from the galactic center and the
distribution of the detectable matter in the galaxy. The procedure
we follow is almost that of Sanders and McGaugh (2002):

In order to calculate the MOND rotation curves, the first step is
determining the Newtonian acceleration of the detectable matter,
$g_N$ via the classical Poisson equation. Given the Newtonian
acceleration, the effective acceleration is calculated from the MOND
Eq. (\ref{gmond}). For MOG, we approximate the galaxy by a
spherically symmetric system. The error committed in this
approximation, as described in Binney \& Tremaine (1987), is on the
order of 15\%.

We assume a constant  $M_*/L$ ratio throughout the galaxy, though
this is not strictly the case, because of the color gradient in
spiral galaxies. However, in seven bulged spirals we find assigning
different $M_*/L$  to the bulge and the disk improves the fit to the
observed data.

We assume the HI gas is in co-planer rotation about the center of
the galaxy, an assumption which may not hold in galaxies with strong
bars (\cite{san02}).

Given the observed distribution of the baryonic matter (stellar and
gaseous disks, plus a spheroidal bulge, if present), the effective
radial gravitational force, and subsequently the circular speed, is
calculated from Eqs. (\ref{gmond}) and  (\ref{gmog}) - (\ref{cdm}).
Fitting of the calculated rotation curves to the observed data
points is achieved by adjusting the $M_*/L$ ratio, through a
least-square $\chi^2$, defined as
\begin{equation}
 \chi^2=\frac{1}{N}\sum_{i=1}^{N}\frac{(v^{i}_{theory}-v_{obs}^{i})^2}{\sigma_i^2}\label{chi},
\end{equation}
\noindent where $\sigma_i$ is the observational uncertainty in the
rotation speeds. The $M_*/L$ ratio of the disk and of the bulge are
our ultimate results.

\section{Numerical results}\label{res}

All models trace the observed data with reasonable detail. The
best-fit $\chi^2$ and $M_*/L$ values are listed in Table \ref{t1}.
Figures \ref{f1} - \ref{f3} show fits of theoretically constructed
rotation curves to the observations of 46 galaxies. The general
trend of HSB curves ( steep rise to a maximum followed by gradual
decline to an almost flat asymptote), and of LSB curves ( slow rise
often with  no asymptote in sight) are evident.

Seven galaxies have prominent bulge components. One expects a bulge
with an older population of stars to have a higher $M_*/L$ ratio
than a disk with a younger population. Therefore, to obtain a better
fit for these galaxies, we have allowed the model to choose
different $M_*/L$s for the disk and the bulge.  The result, shown in
Table \ref{t3}, confirms the expectation. The minimum $\chi^2$s of
Table \ref{t3} are detectably lower than those of the corresponding
entries of Table \ref{t1} obtained by a single $M_*/L$ fit.
Nevertheless, one galaxy in MOND3, NGC 801, and four in MOND1, NGC
801, NGC 5371, UGC 2885, and NGC 5907, predict untenably lower
$M_*/L$ for the bulge than for the disk.

McGaugh (private communication) advises us that this oddity might
partially be owing to the low resolution of HI data and/or the sharp
rise of  $v(r)$ in the bulge.  If $v(r)$ is not quite resolved, one
tends to underestimate it, and the $M_*/L$  of the bulge with it. In
many of these galaxies it is indeed not obvious if the inner
component is really a ``bulge" in the classical sense of a 3D
component with an $r^{1/4}$ profile. It is also said that neglecting
the flattening of a bulge leads to an error in its mass-to-light
ratio (Noordermeer 2008). In NGC 6946, there is a tiny bulge (4\% of
the total light in B-band). It shows up as a small kinematic bump in
the inner 1 kpc in high-resolution Fabry-P\'erot data
(\cite{bla04}). Moreover, the HI distribution in NGC 6946 is not
symmetric in the galactic plane.  It is patchy and seems to deviate
from circular orbits (\cite{car90}).

Ten entries in Table \ref{t1} have unacceptably large $\chi^2$s.
Suspecting the failure of the assumption of constant $ M_*/L$, we
have followed Barnes et al. (2007), and examined a radially varying,
$M_*/L= (M_*/L)_0 + m r$. The best-fit values of the constants $
(M_*/L)_0$ and $m $ for some of them are displayed in Table
\ref{variable}. The slope, $m$, is much too small to result in
appreciably  lower $\chi^2$. The reason for the failure should lie
elsewhere. For example, the assumption of cold unobservable
molecular gas in the galactic disk (Tiret \& Combes 2009) leads to
better fits with lower $\chi^2$.

\section{Color -- $M_*/L$ correlation }\label{ml}

How realistic are the inferred  $M_*/L$ ratios? Stellar population
synthesis (SPS) models predict a linear relation between colors and
$M_*/L$ ratios. Redder galaxies should have higher $M_*/L$ ( see,
e.g., \cite{bel01, bel03, port04}). The slope of this linear
relation does not depend on the exact details of the history of star
formation, i.e. the assumed IMF. But depending on how many stars are
present at the low-mass end of the stellar IMF, the color-$M_*/L$
curve shifts up and down. This is because low-mass stars contribute
significantly to the mass of a population, but not as much to its
luminosity and color (\cite{bel01}).

In the SPS scheme Salpeter's (1955) IMF  overestimates the $M_*/L$
ratios of many of the galaxies and violates the condition of `less
disk mass than the mass of maximum disk'. To remedy this, Bell et
al. (2003) scale down Salpeter's IMF and come up with a limit for
the color-$M_*/L$ relation above which the physical viability is not
guaranteed. Their suggested relation is

\begin{equation}
 \log(M_*/L_B)= 1.74(B-V)-0.94\label{color}.
\end{equation}
There are other IMF that lead to slightly different relations. For
example, based on an analysis of the vertical velocity dispersion of
stars, Bottema (1997) argues for a substantially submaximal $M/L$
ratio for all disk-dominated galaxies.  Alternatively Kroupa (2001)
introduces a turnover at the low-mass end of his IMF.

\begin{figure*}{}
\begin{center}
\resizebox{18.5cm}{!}{\includegraphics[width=120mm,
height=120mm]{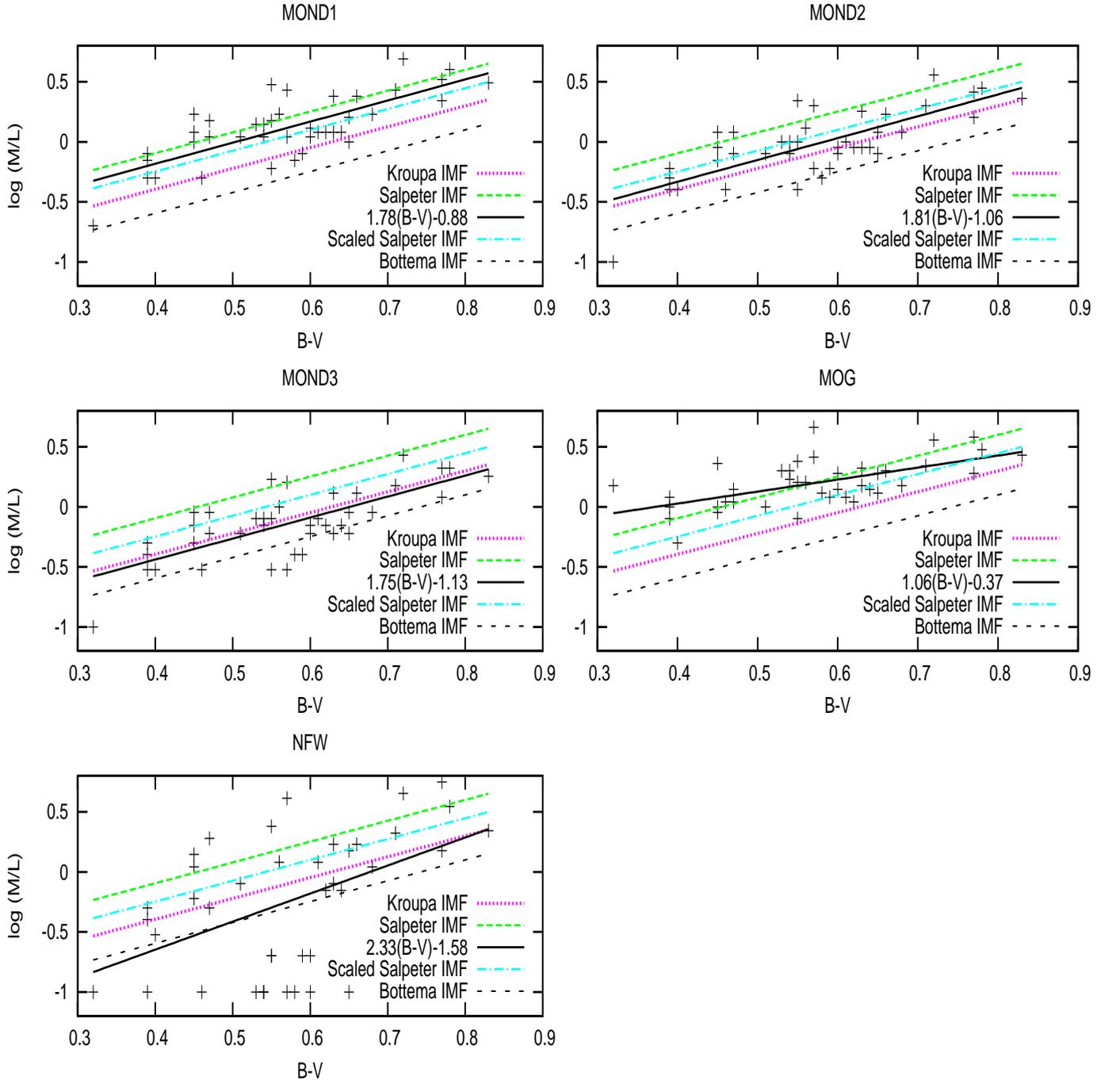}} \caption{ Plots of $ M_*/L$ versus
$B-V$. The solid line in each panel is the best fit to the
prediction of the model in question. The slopes and
\emph{y}-intercepts of best-fitted lines are shown in the panels.
Slopes of MOND1, MOND2, and MOND3 are reasonably close to the
prediction of SPS, 1.74. That of MOG is not. Data points in NFW
panel are much dispersed  to conclude a meaningful correlation.  In
MOND1, MOND2, and MOND3 errors in the slopes, $\pm{(0.23,
0.21,0.26)}$, respectively, are small enough to distinguish one
model from the other. The errors in the \emph{y}-intercepts,
$\pm(0.12, 0.14, 0.15)$, respectively, are small enough to
distinguish one IMF from the other. The other lines are the
theoretical predictions of SPS with different IMFs (\cite{bel01,
bel03}).  They are included here for comparison. They have almost
the same slope, but different \emph{y}-intercepts.
}\label{ML-Color1}
\end{center}
\end{figure*}

In Eq. (\ref{color}) the slope $1.74$ is not sensitive to variations
in IMF, but the $y$-intercept is. To obtain the equivalent relation
for standard Salpeter's, Kroupa's, and Bottema's IMF one should
shift Eq. (\ref{color}) and the plots in Fig. (\ref{ML-Color1}) up
and down by about (0.15, -0.15, -0.35) dex, respectively
(\cite{bel03}).

In Fig. \ref{ML-Color1} we contrast  $M_*/L$ ratios of the three
gravity models against the predictions of SPS, where we use the
B-band luminosities of Sanders \& McGaugh (2002). In each frame the
solid line is the best fit to the data points obtained from the
analysis of the rotation curves. The theoretical SPS predictions of
Bell \& de Jong (2001), and Bell et al. (2003) for different IMFs
are also plotted. The slope of MOND1, $1.78 \pm 0.23$, of MOND2,
$1.81\pm 0.21$, and of MOND3, $1.75\pm 0.26$ are reasonably close to
that of Eq. (\ref{color}). The corresponding \emph{y}-intercepts,
$-0.88\pm0.14$, $-1.06\pm0.12$, and $-1.13\pm0.15$, respectively,
are also in harmony with that of Eq. (\ref{color}). The
uncertainties in slopes and \emph{y}-intercepts are in the
1-$\sigma$ error range. The errors in \emph{y}-intercepts are small
enough to enable one to distinguish between different IMFs. MOND1
falls somewhere between standard Salpeter's and scaled Salpeter's
IMF. MOND2 agrees with Kroupa's and scaled Salpeter's IMF. MOND3 is
in good agreement with Kroupa's IMF.

The slope for MOG, $1.06 \pm 0.21$, cannot be reconciled with SPS
predictions. The case of NFW is also questionable. Although the
slope, $2.33\pm 0.67$, is consistent with 1.74, considering its
large error bar, the dispersion of the simulated data points is too
large to conclude a meaningful color-$M_*/L$ correlation.

Any alternative gravity can have a dark matter equivalent.
Deviations from the Newtonian gravity can be attributed to a
hypothetical dark entity and a dark density profile  calculated
through Poisson's equation, for instance. One feature however
distinguishes this interpretation from the conventional CDM
scenarios. Here, there is a well defined relation between the
baryonic matter and its so-interpreted dark companion. While in CDM
models, baryonic and dark matters may co-exist independently. In our
opinion, the reason for the good agreement of MOND with SPS
predictions and non-compliance of CDM  with it lies in the existence
or non-existence of this relation between the observable and
non-observable matters. In MOND, baryonic matter plays a pivotal
role, and the dark entity owes its existence to it.  This is not the
case in CDM. Dark matter is allowed to play a role independently
ofthe observable matter. As for MOG, we are not in a position to
express an opinion.

Let us summarize our  conclusion: a) the SPS scheme can distinguish
between different gravity models and b) the two together can choose
between different IMFs. The mere fact that a gravity theory
reproduces the observed rotation curves satisfactorily does not tell
the whole story.

\section{Concluding remarks}\label{con}

At least on galactic scales, dynamics of spirals cast doubt on the
viability of the classical theories of gravitation.  A number of
alternative theories  are capable of reproducing the rotation curves
of spirals with acceptable detail, a nontrivial fact that deserves
attention. In this paper we used two alternative theories of
gravitation, MOND and MOG, and a CDM model to deduce the dynamics of
a well-sized sample of high- and low-surface brightness galaxy
types, and checked the results against observations. The models are
not equivalent although they all simulate the rotation curves in
more or less to the same degree of accuracy, .

In MOND and MOG, rotation curves are constructed with only one free
adjustable parameter, the stellar mass-to-light ratio. This is in
contrast to the CDM model, where an additional parameter is needed
to describe the dark component.

There are cases of bulged galaxies where fits to observations lead
to lower $M_*/L$ ratios for the bulge than for the disk. This might
be owing to the low resolution of the HI data and to the inner
$v(r)$, and/or uncertain size of the bulge.

Stellar population synthesis models impose constraints on $M_*/L$.
Redder galaxies should have higher $M_*/L$ ratios. MOND with
different interpolating functions meets this expectation, albeit
with different IMFs. This is remarkable, because there is no
explicit/implicit connection between the basic tenets of the SPS and
MOND. On the other hand, MOG does not meet the SPS constraints, and
the data points  of CDM show a large scatter, preventing one to
conclude a meaningful color-$M_*/L$ correlation.

The SPS predictions of $M_*/L$ ratios are sensitive to the adopted
IMF. The $M_*/L$ ratios inferred from Salpeter's IMF are notably
higher than those obtained from Kroupa's. MOND3 favors Kroupa's IMF.
It produces lower $M_*/L$ and implies lesser disk masses.

\begin{acknowledgements}

Stacy McGaugh provided us with his recent data on rotation curves.
We thank him and also Eric Bell, Roelf de Jong, Mohammad Dehghan
Niry, and Pavel Kroupa for their useful comments. We sincerely thank
Prof. Dr. Yousef Sobouti, the director of IASBS, for useful comments
and important contribution in writing this article.
\end{acknowledgements}


\begin{table*}
\begin{center}
\begin{tabular}{ccccccccccccccc}
\hline

Galaxy (Type) & B-V & $M^*/L$&$\chi^{2}$ & $M^*/L$&$\chi^{2}$& $M^*/L$&$\chi^{2}$&$M^*/L$&$\chi^{2}$&$M^*/L$&$\chi^{2}$\\
&&Mond1&Mond1&Mond2&Mond2&Mond3&Mond3&MOG&MOG&NFW&NFW\\

\hline &&&&&&&&&&&\\
HSB Galaxies&&&& &&&&&&&\\
\hline
      M  33     (Sc) ~~~~~~  &0.55  &0.6  &30.22  &0.4   &36.47 &0.3  &39.25  &0.8  &45.03  &0.2&  125.00\\
      NGC 300   (Sc)~~       &0.58  &0.7  & 2.26  &0.5   &2.51  &0.4  & 2.35  &1.3  &2.54   &0.1&     2.94\\
      NGC 2903  (Sc)         &0.55  & 3.0 & 6.07  &2.2   &5.72  &1.7  & 4.79  &2.4  &7.18   &2.4&     3.60\\
      ~~NGC 3726  (SBc)      &0.45  & 1.0 & 3.47  &0.7   &3.86  &0.5  & 4.60  &0.9  &5.00   &0.6&  3.03\\
      ~~NGC 3769  (SBb)      &0.64  & 1.2  & 0.75 &0.9  &0.65  &0.7  & 0.70   &1.4  &1.06   &0.7&     0.42\\
      NGC 3877  (Sc)         &0.68  & 1.7  & 2.66 &1.2  &2.62   &0.9  & 2.60 &1.5  &3.07   &1.1&     2.51\\
      NGC 3893  (Sc)         &0.56  & 1.7  & 4.12 &1.3  &2.11   &1.0  & 1.88  &1.6  &2.76   &1.2&     1.89\\
      ~~NGC 3949  (Sbc)      &0.39  & 0.8  & 5.34 &0.6  &3.88  &0.5  & 3.97   &0.8  &5.00   &0.5&  3.29\\
      ~~~~NGC 3953  (SBbc)   &0.71  & 2.7  & 1.13 &2.0  &0.48  &1.5  & 0.47   &2.2  &0.44   &2.1&     1.05\\
      ~NGC 3972  (Sbc)       &0.55  & 1.5  & 3.22 &1.0  &3.03  &0.8  & 3.08   &1.6  &2.85   &0.2&     1.25\\
      ~~~~NGC 3992  (SBbc)   &0.72  & 4.9  & 0.65 &3.6  &0.88  &2.7  & 1.02   &3.6  &2.08   &4.5&     4.11\\
      NGC 4013  (sb)         &0.83  & 3.1  & 1.37 &2.3  &1.62  &1.8  & 2.24   &2.7  &2.05   &2.2&     1.63\\
      ~~~~NGC 4051  (SBbc)   &0.62  & 1.2  & 0.88 &0.9  &0.78  &0.7  & 0.90   &1.1  &0.76   &0.7&     0.78\\
      NGC 4085  (Sc)         &0.47  & 1.1  & 6.84 &0.8  &6.02  &0.6  & 6.03    &1.1  &6.94   &0.5&     3.93\\
      ~~NGC 4088  (Sbc)      &0.51  & 1.1  & 1.49 &0.8  &1.62  &0.6  & 1.75    &1.0  &1.97   &0.8&     1.24\\
      ~~NGC 4100  (Sbc)      &0.63  & 2.4  & 2.07 &1.7  &1.76  &1.3  & 1.58   &2.0  &1.84   &1.9&     1.69\\
      NGC 4138  (Sa)         &-     & 3.5  & 1.61 &2.7  &0.98  &2.0  & 0.81   &3.2  &1.20   &2.9&    1.09\\
      NGC 4157  (Sb)         &0.66  & 2.4  & 0.92 &1.7  &0.85  &1.3  & 0.87   &2.0  &0.84   &1.7&     0.89\\
      NGC 4217  (Sb)         &0.77  & 2.2  & 3.95 &1.6  &2.92  &1.2  & 2.90   &1.9  &2.63   &1.5&     3.02\\
      ~~~NGC 4389  (SBbc)    &-     & 0.4  & 5.36 &0.3  &5.42  &0.2  & 5.57   &0.6  &6.33   &0.1&     4.33\\
      ~~~NGC 5585  (SBcd)    &0.46  &0.5   &10.43 &0.4  &10.22 &0.3  & 10.46  &1.1  &16.87  &0.1&      19.03\\
      ~~~~~NGC 6946  (SABcd) &0.40  &0.5   &11.46 &0.4  &17.90 &0.3  & 21.61  &0.5  &31.75  &0.3&     1.01\\
      NGC 7793  (Scd)        &0.63  &1.2   &1.48  &0.9  &1.44  &0.6  & 1.48   &1.5  &1.02   &0.8&     1.91\\
      UGC 6399 (Sm)          &-     &1.0   &0.16  &0.8  &0.17  &0.6  & 0.10   &1.8  &0.04   &0.1&     1.48\\
      UGC 6973 (Sab)         &-     &2.7   &20.46 &2.2  &11.81 &1.7  & 10.57  &2.6  &20.24  &2.0&     6.48\\
       NGC 801$^b$ (Sc)~~    &0.61  &1.2   &23.14 &1.0  &14.75 &0.8  & 14.15  &1.2  &23.90  &1.2&     49.61\\
       ~~NGC 2998$^b$  (SBc) &0.45  &1.2   &2.64  &0.9  &2.43  &0.7  & 2.96   &1.0  &2.35   &1.4&    6.80\\
       ~~~~NGC 5371$^b$ (S(B)b)&0.65&1.6   &10.02 &1.2  &8.32  &0.9  & 6.93   &1.3  &6.65   &1.5&     7.65\\
       ~NGC 5533$^b$ (Sab)   &0.77 &3.3   &2.33  &2.6  &1.61   &2.1   & 1.11  &3.8  &8.50   &5.6&     19.61\\
       ~~NGC 5907$^b$ (Sc)~~ &0.78 &4.0   &2.93  &2.8  &3.82   &2.1   & 4.27  &3.0  &6.10   &3.5&     11.23\\
       ~~NGC 6674$^b$ (SBb)  &0.57 &2.7   &10.96 &2.0  &7.65   &1.6   & 6.64  &2.6  &41.06  &4.1&      66.96\\
       ~UGC 2885$^b$ (Sbc)   &0.47 &1.5   &2.80  &1.2  &2.98   &0.9   & 3.04  &1.4  &6.64   &1.9&      14.85\\
       \hline&&&&&&&&&&&\\
       LSB Galaxies&&&&  &&&&&\\
       \hline
       DDO 168   (SO)~~      &0.32  &0.2   &11.50  &0.1  &14.35 & 0.1 &21.56 &1.5  &14.64  &0.1&       26.67\\
       NGC 247   (SBc)~      &0.54  &1.1  &3.71  &0.8  &3.91    & 0.7  & 4.16  &2.0  &3.74   &0.1&       10.34\\
       NGC 1560  (Sd)~       &0.57  &1.1  &3.35  &0.6  &1.94    & 0.3  & 1.52  &4.6  &10.56  &0.1&      17.00\\
       NGC 3917  (Scd)       &0.60  &1.3  &4.49  &0.9  &4.58    & 0.7  & 4.58 &1.4  &4.03   &0.2&      6.51\\
       NGC 4010  (SBd)       &0.54  &1.4  &1.81  &1.0  &1.74    & 0.8  & 1.76  &1.7  &1.16   &0.1&      2.42\\
       NGC 4183  (Sa)~~      &0.39  &0.7  &0.98  &0.5  &0.98    &0.4   & 1.09  &1.0  &1.54   &0.4&      0.20\\
       UGC 128   (Sdm)~      &0.60  &1.1  &0.48  &0.8  &0.54    &0.6   & 0.63  &1.9  &0.36   &0.1&      2.57\\
       UGC 6446  (Sd)~~      &0.39  &0.5  &2.30  &0.4  &3.29    &0.3   & 4.49  &1.2  &2.35   &0.1&      0.30\\
       UGC 6667  (Scd)       &0.65  &1.0  &0.94  &0.8  &0.88    &0.6   & 0.69  &1.9  &0.59   &0.1&      3.95\\
       UGC 6917  (SBd)       &0.53  &1.4  &0.64  &1.0  &0.69    &0.8   & 0.72  &2.0  &0.84   &0.1&      0.49\\
       UGC 6923  (Sdm)       &-     &0.8  &1.17  &0.6  &1.16    &0.4   & 1.03  &1.4  &2.28   &0.1&      0.56\\
       UGC 6930  (SBd)       &0.59  &0.8  &0.28  &0.6  &0.34    &0.4   & 0.54  &1.2  &0.34   &0.2&      0.19\\
       ~UGC 6983  (SBcd)     &0.45  &1.7  &1.30  &1.2  &1.46    &0.9   & 1.68  &2.3  &1.90   &1.1&      0.54\\
       UGC 7089  (Sdm)       &-     &0.2  &0.14  &0.2  &0.40    &0.1   & 0.25  &0.6  &0.11   &0.9&      3.18\\
      \hline
\end{tabular}
\end{center}
\caption{ Best-fit reduced $\chi^2$ and $M_*/L$ values of 32 HSB and
14 LSB galaxies in MOND with different interpolating functions, MOG,
and NFW models. For explanation of models see text. Hubble types are
from NASA/IPAC Extragalactic Database (NED). Bulged galaxies are
marked by a superscript '${^b}$'.\label{t1}}
\end{table*}


\begin{table*}
\begin{center}
\begin{tabular}{cccccccc}
 Galaxy          &$\chi^{2}_{MOND3}$
&$(\frac{M^*}{L})^{disk}_{MOND3}$ &$(\frac{M^*}{L})^{bulge}_{MOND3}$
&$\chi^{2}_{MOND1}$ &$(\frac{M^*}{L})
^{disk}_{MOND1}$ &$(\frac{M^*}{L})^{bulge}_{MOND1}$   \\
\hline
      NGC 801  &12.69 & 1.1$^*$ &  0.7    &17.72  &2.2$^*$   &1.1  \\
      NGC 2998 & 2.22 & 0.6   &  0.8    & 2.50  &1.2     &1.3  \\
      NGC 5371 & 6.34 & 0.8   &  1.0    & 9.96  &1.7$^*$   &1.6  \\
      NGC 5533 & 0.98 & 0.7   &  2.2    & 2.12  &0.1     &3.7  \\
      NGC 5907 & 3.5  & 1.8   &  3.2    & 2.80  &4.1$^*$   &3.6  \\
      NGC 6674 & 6.32 & 0.4   &  1.8    &10.83  &1.5     &2.9  \\
      UGC 2885 & 2.07 & 0.8   &  1.1    & 2.80  &1.5$^*$     &1.4  \\
\hline
\end{tabular}
\end{center}
\caption{ Same as Table 1 for seven bulged galaxies. Disks and
bulges have different $M_*/L$ ratios. One galaxy in MOND3 model and
four in MOND1's  predict untenably lower $M_*/L$ for the bulge than
for the disk. These are marked by asterisks in the $M_*/L$ columns.
\label{t3}}
\end{table*}


\begin{table*}
\begin{center}
\begin{tabular}{cccccccc}

Galaxy          &$\chi^{2}_{MOND3}$ &$(\frac{M^*}{L})_{0,MOND3}$ &$m_{MOND3}$ & $\chi^{2}_{MOND1}$ &$(\frac{M^*}{L})_{0,MOND1}$ & $m_{MOND1}$  \\
\hline  M 33       & 37.4   &0.4  &-0.01    &27    &0.6    &0.01  \\
      DDO 168      & 20.19  &0.1   &-0.02    &10    &0.1    &0.07  \\
      NGC 801      & 14.15  &0.8   & 0.00    &19.88 &1.1    &0.01  \\
      \hline
\end{tabular}
\end{center}
\caption{ Linear, $M_*/L= (M_*/L)_0 + mr$, fit for 3 cases not well
explained by either MOND3 formalism or by that of  MOND1. A
comparison with constant $M_*/L$ of these galaxies (Table 1) shows
no significant improvement. \label{variable}}
\end{table*}


\end{document}